\documentclass{article}
\newif\ifpaperanonymous
\paperanonymousfalse

\usepackage[T1]{fontenc}
\usepackage{microtype}
\usepackage{graphicx}
\usepackage{booktabs}
\usepackage{array}
\usepackage{xurl}
\usepackage{xcolor}
\usepackage{hyperref}
\usepackage[accepted]{mlsys2025}

\usepackage{etoolbox}
\usepackage{balance}
\usepackage{placeins}
\makeatletter
\renewcommand{\Notice@String}{}
\let\paper@template@makecaption\@makecaption
\long\def\@makecaption#1#2{%
  \ifdefstring{\@captype}{figure}{%
    \begingroup
    \normalfont\normalsize
    \let\small\normalsize
    \let\footnotesize\normalsize
    \let\sl\upshape
    \paper@template@makecaption{#1}{#2}\par
    \endgroup
  }{\paper@template@makecaption{#1}{#2}}}
\makeatother
\fussy
\newcolumntype{L}[1]{>{\raggedright\arraybackslash}p{#1}}
\newcommand{\sys}{UlaCache}
\newenvironment{paperintroduction}{%
  \begingroup
}{\par\endgroup}
\newcommand{\paperauthor}[2]{%
  \parbox[t]{.25\textwidth}{\centering
    {\fontsize{13}{15}\selectfont #1\par}%
    \vspace{1pt}%
    {\normalsize\rmfamily #2\par}}}
\newcommand{\papercorrespondence}{%
  \begingroup
  \renewcommand{\thefootnote}{\fnsymbol{footnote}}%
  \footnotetext[1]{Corresponding author.}%
  \endgroup}
\makeatletter
\newcommand{\resultcaption}[2]{%
  \begingroup\def\@captype{figure}\caption{#1}\label{#2}\endgroup}
\makeatother
\mlsystitlerunning{UlaCache: Unified Logical Addressing for Sparse LLM Serving}
\hypersetup{hypertexnames=false,pdfview=XYZ,
  pdftitle={UlaCache: Unified Logical Addressing for Sparse LLM Serving}}

\hypersetup{
  pdfauthor={Isaac Zhu, Hugh Yin, Orion Wang, Alan Cao, Rocky Wang, Hscos Zhang, Jack Li, Jason Zhao},
  pdfsubject={arXiv preprint}}

\usepackage{float}
\begin{document}
\twocolumn[
\mlsystitle{UlaCache: Unified Logical Addressing\\for Sparse LLM Serving}

\begingroup
\setlength{\topsep}{0pt}
\setlength{\parskip}{0pt}
\begin{center}
\makebox[.82\textwidth]{%
  \paperauthor{Isaac Zhu}{isaac@scitix.ai}\hfill
  \paperauthor{Hugh Yin}{hugh@scitix.ai}\hfill
  \paperauthor{Orion Wang}{orion.wang@scitix.ai}}\par
\vspace{10pt}
\makebox[.82\textwidth]{%
  \paperauthor{Alan Cao}{sage@scitix.ai}\hfill
  \paperauthor{Rocky Wang}{rocky@scitix.ai}\hfill
  \paperauthor{Hscos Zhang}{cos120@scitix.ai}}\par
\vspace{10pt}
\makebox[.82\textwidth]{%
  \paperauthor{Jack Li}{jack@scitix.ai}%
  \hspace{.035\textwidth}%
  \paperauthor{Jason Zhao\textsuperscript{*}}{jason@scitix.ai}}\par
\vspace{16pt}
{\Large Scitix.AI\par}
\vspace{14pt}
\end{center}
\endgroup

\mlsyskeywords{LLM serving, unified KV addressing, sparse attention, GPU--CPU memory, tensor parallelism, data parallelism}
\begin{abstract}
Long-context inference in large language models (LLMs) imposes substantial memory requirements for the key--value (KV) cache. Token-selective sparse attention reduces attention computation and memory traffic, but KV entries must still be retained for future queries. KV storage capacity therefore remains a constraint on decoding concurrency. Combining GPU and CPU memory expands this capacity but requires efficient access across heterogeneous memory pools. We present \sys, a KV cache management system that provides unified logical addressing across GPU and CPU memory. Its logical-slot interface enables the inference engine to allocate, write, access, and release KV entries using stable identifiers across attention parallelism modes. \sys\ manages physical placement, residency, and transfers, updating address mappings after KV transfers. Parallel Bucketed LRU (PB-LRU) reduces the overhead of KV replacement by parallelizing recency updates and eviction-candidate selection within a GPU memory pool shared across requests. We evaluate \sys\ against state-of-the-art baselines. Under tensor-parallel attention (TP), \sys\ provides up to $11.4\times$ the usable KV cache capacity of HiSparse. For GLM-5.3 on eight H200 GPUs, \sys\ achieves peak decoding throughput of $1.10\times$ and $1.59\times$ that of HiSparse under data-parallel attention (DPA) and TP, respectively. For HY4 Preview on eight B300 GPUs, the corresponding throughput ratios are $1.27\times$ and $1.93\times$.

\end{abstract}
]

\papercorrespondence

\hypersetup{pdftitle={UlaCache: Unified Logical Addressing for Sparse LLM Serving}}

\begin{paperintroduction}
\section{Introduction}
\label{sec:introduction}

% Reserve the first column for the Introduction; place the teaser atop the next column.
\suppressfloats[t]
\begin{figure}[t]
\centering
\includegraphics[width=\columnwidth,trim=0 12.4163bp 0 0bp,clip]{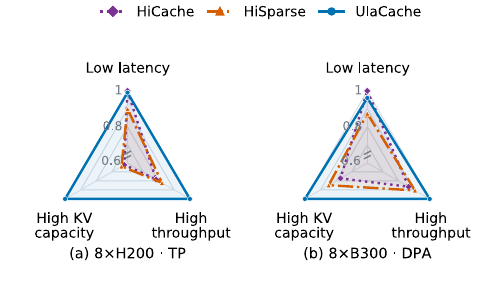}
\vspace{-10pt}
\caption{TP on $8\times$H200 and DPA on $8\times$B300. Normalized per panel; latency axis: 0.6--1.0, others: 0--1.}
\vspace{-10pt}
\label{fig:performance-overview}
\end{figure}

Long-context LLM serving needs enough KV memory to keep many requests active. Batching amortizes model execution~\cite{pagedattention,sarathi}, but KV storage grows with context length and concurrency, limiting the decode batch. Increasing effective KV capacity within a fixed memory budget is therefore central to throughput scaling.

Sparse attention reduces KV reads by selecting tokens or blocks~\cite{dsa}. DeepSeek-V4.1-Flash~\cite{deepseekv41}, the GLM-5 family~\cite{glm5}, Qwen3.8-Flash-Next~\cite{qwen38}, Hy4 Preview~\cite{hy4preview}, and Step 5 Preview~\cite{step5preview} adopt this approach for long contexts. Since index-based selections change across steps, unselected KV must remain available for future queries~\cite{dsa,hisparse}. The storage layer must therefore retain the full KV cache while serving sparse, position-addressed reads; reading less KV per step does not remove the capacity constraint.

CPU memory can extend KV storage, but its effect depends on how it is used. HiCache and LMCache restore reusable prefixes into the inference engine's GPU slots, saving prefill computation while leaving active decode capacity bounded by that allocation~\cite{hicache,lmcache}. HiSparse instead stores complete KV caches for active requests in CPU memory and fetches selected KV into GPU caches~\cite{hisparse}. This supports decoding beyond GPU KV capacity, but the cached copies add no capacity for distinct KV records.

We present \sys, a KV storage system for sparse attention. It provides one logical address space across GPU and CPU memory and manages KV placement and transfers for the inference engine. Prefill writes KV into this storage, and decoding reads and extends it. The interface supports both colocated and disaggregated prefill/decode deployments. Our optimizations target KV capacity and sparse-access overhead during decoding, while retaining the engine's prefill computation. Realizing this design poses three challenges.

\textbf{Challenge 1: Unified KV addressing and storage organization.} Aggregating GPU and CPU memory requires both tiers to provide storage for distinct KV records. Selected CPU-resident KV must be transferred to the GPU for attention. When the GPU pool is full, this also requires moving displaced KV to CPU memory. These migrations change physical locations but must preserve the inference engine's logical references. The challenge is to organize KV storage and transfers across both tiers under unified logical addressing, while reducing redundant storage and unnecessary data movement.

\textbf{Challenge 2: Adapting logical KV addressing to attention parallelism.} Beyond GPU--CPU storage organization, attention parallelism determines how KV is placed and accessed across ranks. Data-parallel attention (DPA) keeps each request's KV on its serving rank. Tensor parallelism partitions KV heads across ranks, with replication when there are fewer KV heads than ranks or under multi-head latent attention (MLA)~\cite{mla,helix}. Context parallelism (CP) partitions KV by token position. These layouts determine which ranks a sparse read must reach and which KV shard or replica it accesses. The challenge is to adapt address translation and KV access paths to each parallelism strategy's KV layout, while providing a unified logical address space to the inference engine.

\textbf{Challenge 3: Efficient LRU replacement and mapping management.} KV accesses in sparse attention exhibit temporal locality across decoding steps: recently accessed entries tend to be reused~\cite{hisparse}. This makes Least Recently Used (LRU) replacement a suitable policy for managing the GPU cache. LRU requires recency updates on accesses and evictee selection when the cache is full. KV migration additionally requires address mappings to be updated after transfers complete. These operations recur across requests and layers at every decoding step. Serial recency updates or full-cache scans for evictee selection can make management costly as cache capacity and concurrency grow. The challenge is to reduce the overhead of LRU replacement and mapping updates on the decoding critical path.

We make four contributions:
\begin{itemize}
  \item \textbf{Unified GPU--CPU KV storage.} We combine GPU and CPU memory into one logical KV address space, with both tiers storing different KV to increase effective capacity. The inference engine can keep using the same KV references as data moves between the tiers.
  \item \textbf{KV access for different attention parallelism strategies.} We design KV access paths for CP, DPA, and TP according to how each strategy distributes KV and attention computation. The engine accesses KV through a common interface.
  \item \textbf{Parallel GPU cache management.} We share GPU cache capacity across requests so active requests can use space left idle by others. To manage this larger cache efficiently, we group KV by recent use and update the groups in parallel, reducing LRU management overhead.
  \item \textbf{A common KV interface for inference engines.} We expose these capabilities through an interface that works with engine features such as speculative decoding, prefix caching, and CUDA Graphs. The engine retains control over request scheduling and KV reuse, while \sys\ manages storage and transfers. We integrate the interface into SGLang and evaluate DPA and TP on two models and two GPU platforms.
\end{itemize}

Figure~\ref{fig:performance-overview} summarizes our results. In TP, \sys\ provides $9.3$--$11.4\times$ HiSparse's KV capacity under matched memory budgets, with higher decoding throughput and lower TPOT at matched concurrency. Our code is available.\footnote{\url{https://github.com/scitix/UlaCache}}

\end{paperintroduction}

\section{Background and Motivation}
\label{sec:motivation}

\subsection{Background}
\label{sec:background}

\subsubsection{Sparse Attention and KV Cache}

LLM inference consists of prefill and decoding. Prefill computes KV states for the input sequence; decoding adds new states and repeatedly reads the KV cache to generate tokens. The cache avoids recomputation and must remain available while a request is active. Longer contexts and more concurrent requests therefore increase KV storage requirements and can limit decode concurrency~\cite{pagedattention}.

In index-based sparse attention, an indexer selects token or block positions, and the attention operator reads the corresponding KV data~\cite{dsa,glm5}. For a KV cache with $L$ entries and a selection of $k$ positions, the main attention operation reads only the selected $k$ entries, while storage must keep the remaining KV available. The indexer maintains compact selection state separately from the KV records consumed by attention. Because selections change with the query, an entry outside the current working set can be needed at a later step~\cite{hisparse}.

\subsubsection{Attention Parallelism and KV Placement}

Attention parallelism determines which GPUs compute a request and store its KV. TP partitions attention and KV heads across ranks, replicating KV when too few KV heads are available. Multi-head Latent Attention (MLA) shares one compressed KV representation across heads, so conventional head-partitioned TP replicates it across ranks~\cite{mla,helix}.

Under DPA, ranks serve separate requests and store their KV locally. Each request is limited to its rank's capacity unless additional placement and access mechanisms make other ranks' free slots accessible.

Context parallelism (CP) partitions a request's KV by token position across ranks. During decoding, ranks compute attention over local KV shards and combine partial outputs with their softmax normalization statistics, distributing both storage and attention work~\cite{helix}. Other model components can use different partitioning strategies.

\subsubsection{CPU Storage for KV Cache}

CPU memory can retain KV for reuse across requests or back KV for active requests. Hierarchical prefix caches such as HiCache and LMCache keep KV in CPU memory and external storage. A prefix hit restores cached KV into the inference engine's GPU allocation, avoiding repeated prefill computation and improving request-serving efficiency. Transfers can be pipelined with layer execution to reduce exposed loading latency~\cite{hicache,lmcache}. The backing cache stores reusable contexts, while the inference engine manages the GPU slots required for execution.

Sparse KV offloading uses CPU memory to retain KV for active requests~\cite{infinigen,hisparse}. In HiSparse, the full KV cache resides in CPU memory, while GPU buffers cache selected entries across decoding steps~\cite{hisparse}. A selected entry can be read locally on a cache hit; otherwise it must be fetched from CPU memory before attention.

\subsection{Motivation}

% Model formats and budgets follow evaluation/overall/manifest.json and low_latency_manifest.json.
\begin{table}[t]
\caption{Platforms and per-GPU memory budgets (GB).}
\label{tab:platforms}
\centering
\setlength{\tabcolsep}{3pt}
\begin{tabular}{@{}lcc@{}}
\toprule
Configuration & H200 & B300 \\
\midrule
Model & GLM-5.3 NVFP4 & HY4 Preview FP8 \\
GPU: model + KV & 133 & 272 \\
CPU: KV & 128 & 192 \\
\bottomrule
\end{tabular}
\end{table}

\begin{figure}[t]
\centering
\includegraphics[width=\columnwidth,trim=0 6.1504bp 0 1.75bp,clip]{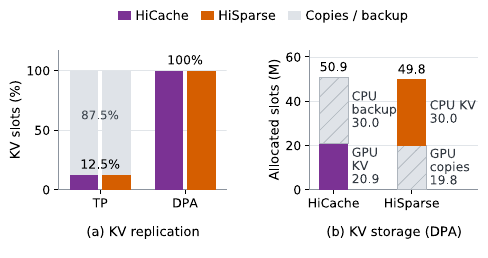}
\vspace{-10pt}
\caption{HY4 Preview on $8\times$B300 (Spec. off): (a) KV replication in TP and DPA; (b) GPU--CPU slot allocation in DPA. Settings: Table~\ref{tab:platforms}.}
\vspace{-10pt}
\label{fig:motivation-history}
\end{figure}

\subsubsection{KV Address Space and Decode Concurrency}

The KV address space available to active requests determines how many distinct KV records can be retained for decoding. We count its capacity in logical slots; replicated or cached copies do not add slots. CPU memory can expand this space, but prefix reuse and storage for active KV have different effects on decode concurrency. The measurements below use the B300 configuration and per-GPU memory budgets in Table~\ref{tab:platforms}.

HiCache and LMCache load reusable prefixes from CPU memory into GPU slots to avoid repeated prefill computation~\cite{hicache,lmcache}. Figure~\ref{fig:motivation-history}(b) shows HiCache's $20.9$ million GPU slots and $30.0$ million CPU backup slots. Active KV must occupy GPU slots during execution, so these $50.9$ million physical slots do not form an aggregate address space for active decoding.

HiSparse retains full KV caches in CPU memory and fetches selected entries that miss its GPU cache~\cite{hisparse}. In Figure~\ref{fig:motivation-history}(b), its $30.0$ million CPU slots and $19.8$ million GPU cache slots total $49.8$ million physical slots. The GPU entries are copies, so only the $30.0$ million CPU slots contribute logical capacity.

In MLA TP, ranks compute different attention heads while storing the same KV representation for each request~\cite{mla,helix}. Figure~\ref{fig:motivation-history}(a) shows that under eight-rank TP, only $12.5\%$ of the HiCache and HiSparse storage slots correspond to distinct KV addresses; the remaining $87.5\%$ holds replicas. DPA removes this cross-rank replication. These counts exclude GPU cache buffers.

These results motivate reducing duplicate KV storage so that GPU and CPU pools, and the ranks participating in TP, can retain more distinct KV under the same memory budget. This separates KV placement from attention computation: a selected token's KV may reside in CPU memory or on another rank rather than on the GPU executing attention. A unified logical address space lets the inference engine identify the selected KV independently of these locations. The storage layer resolves each address and provides the data for attention, updating the mapping after migration while preserving the engine's references.

\subsubsection{Fixed Per-Request Cache Quotas}

Systems such as HiSparse cache selected KV on GPUs across decoding steps~\cite{hisparse}. Hits reuse the cached data, while misses fetch KV from CPU memory. Assigning each request a fixed cache quota bounds its GPU memory use and leaves room for others. With fewer active requests, however, these quotas prevent the remaining requests from using the idle capacity.

Even at batch size 1 (BS1), a fixed per-request buffer can prevent a request from using idle GPU cache space. If the request's KV cache exceeds that buffer, selected entries outside it still require CPU-to-GPU transfers. A shared GPU cache lets active requests use this idle capacity.

Sparse KV accesses exhibit temporal locality~\cite{hisparse}, making LRU a suitable policy. Serial recency updates limit parallel processing across requests, while scanning the full cache for evictees becomes costly as the shared pool grows. The replacement mechanism must therefore update recency and select evictees in parallel while preserving the cache's hit rate.

We propose \sys\ to address these limitations. Table~\ref{tab:kv-management} compares its design with HiCache and HiSparse.

\begin{table}[H]
\caption{KV storage and cache management.}
\label{tab:kv-management}
\centering
\small
\setlength{\tabcolsep}{3pt}
\renewcommand{\arraystretch}{1.08}
\begin{tabular}{@{}>{\raggedright\arraybackslash}m{.35\columnwidth}*{3}{>{\centering\arraybackslash}m{\dimexpr.216666\columnwidth-2\tabcolsep\relax}}@{}}
\toprule
\textbf{Property} & \textbf{HiCache} & \textbf{HiSparse} & \textbf{\sys} \\
\midrule
KV capacity limit & GPU & CPU & GPU+CPU \\
\specialrule{.3pt}{2pt}{2pt}
MLA TP KV layout & Replicated & Replicated & Partitioned \\
\specialrule{.3pt}{2pt}{2pt}
GPU cache policy & \shortstack{Shared\\pool} & \shortstack{Fixed per\\request} & \shortstack{Shared\\pool} \\
\specialrule{.3pt}{2pt}{2pt}
\shortstack[l]{Sparse-access\\replacement} & \shortstack{None\\(resident)} & LRU & PB-LRU \\
\bottomrule
\end{tabular}
\end{table}

\section{System Overview}
\label{sec:overview}

\begin{figure}[!hb]
\centering
\includegraphics[width=\columnwidth,trim=0 1.1674bp 0 0bp,clip]{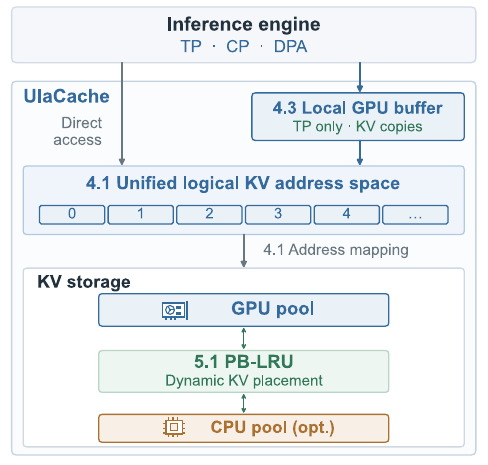}
\vspace{-10pt}
\caption{\sys\ overview: unified KV addressing and storage management.}
\vspace{-10pt}
\label{fig:overview}
\end{figure}

Figure~\ref{fig:overview} shows how \sys\ manages KV through logical addresses, represented as integer IDs that remain valid across physical relocation. Prefill writes KV through allocated IDs. Sparse attention reads the selected KV from the storage pools, with TP ranks caching copies in local GPU buffers.

% Queue the next section's address-space diagram near its discussion.
\begin{figure*}[t]
\centering
\includegraphics[width=\textwidth,trim=0 3.2bp 0 0bp,clip]{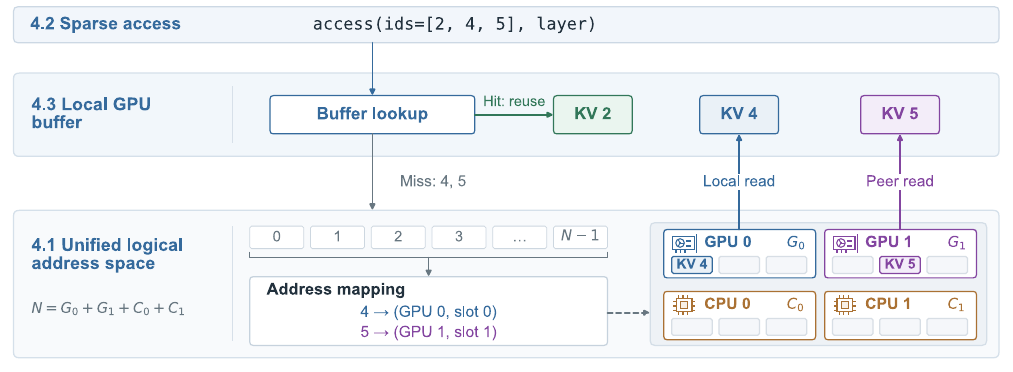}
\vspace{-10pt}
\caption{TP KV access: buffer hits bypass address mapping; misses map and read KV.}
\vspace{-10pt}
\label{fig:gpu-pooling}
\end{figure*}

\noindent\textbf{Unified logical address space.} Address mappings translate logical addresses to physical GPU or CPU slots (Section~\ref{sec:memory}). CP and DPA cover each rank's own pools; TP aggregates the participating ranks' pools. KV resides in these pools, while TP computing ranks cache selected copies in local GPU buffers to reuse hits.

\noindent\textbf{GPU--CPU scheduling with PB-LRU.} PB-LRU uses recent accesses to decide which KV stays on the GPU (Section~\ref{sec:residency}). Parallel GPU operations update access information and select replacement slots. \sys\ transfers the selected CPU KV in and displaced GPU KV out, then updates their address mappings.

\noindent\textbf{Logical KV memory interface.} The inference engine controls logical allocation, reuse, and release; \sys\ provides physical backing and prepares selected KV for attention (Section~\ref{sec:api}). Access returns GPU storage and physical slot indices, with stream dependencies ordering transfers before attention reads.

\section{Unified KV Storage and Access}
\label{sec:memory}

\sys\ supports GPU pools with optional CPU pools in all supported parallelism modes. CP and DPA already partition KV without replication, so \sys\ manages each rank's local storage. TP layouts that replicate KV require a different access path: \sys\ partitions storage across participating ranks and gathers selected KV into local buffers for attention. Figure~\ref{fig:gpu-pooling} shows the TP access path: local buffer hits reuse KV directly, while misses are mapped to physical slots and fetched into the buffer from the storage pools.

\subsection{Unified Logical Address Space}
\label{sec:logical-space}

Storage rank $r$ contributes $G_r$ GPU slots and optionally $C_r$ CPU slots. In CP and DPA, \sys\ handles KV placement, address resolution, and migration within each rank's independent $G_r+C_r$ slots. In TP, $P$ ranks store distinct KV, yielding GPU capacity $\sum_{r=0}^{P-1}G_r$ and total logical capacity $N=\sum_{r=0}^{P-1}(G_r+C_r)$. Each slot holds a token's complete KV representation for one layer, including its compressed latent vector for MLA. Layers sharing sparse selections retain their own KV data but share address mappings and replacement decisions; promotion or eviction applies across these layers.

The ordinary TP MLA layout assigns logical address $t$ to storage rank $t \bmod P$ and rank-local logical position $\lfloor t/P\rfloor$. Without CPU storage, this directly identifies the GPU slot; otherwise, the address mapping resolves the current GPU or CPU location. GPU pool sizing accounts for weights, buffers, Index storage, metadata, and workspace; cached copies add no logical KV capacity.

% Queue the following PB-LRU section's diagram before its discussion.
\begin{figure*}[t]
\centering
\includegraphics[width=\textwidth,trim=0 -0.375bp 0 0bp,clip]{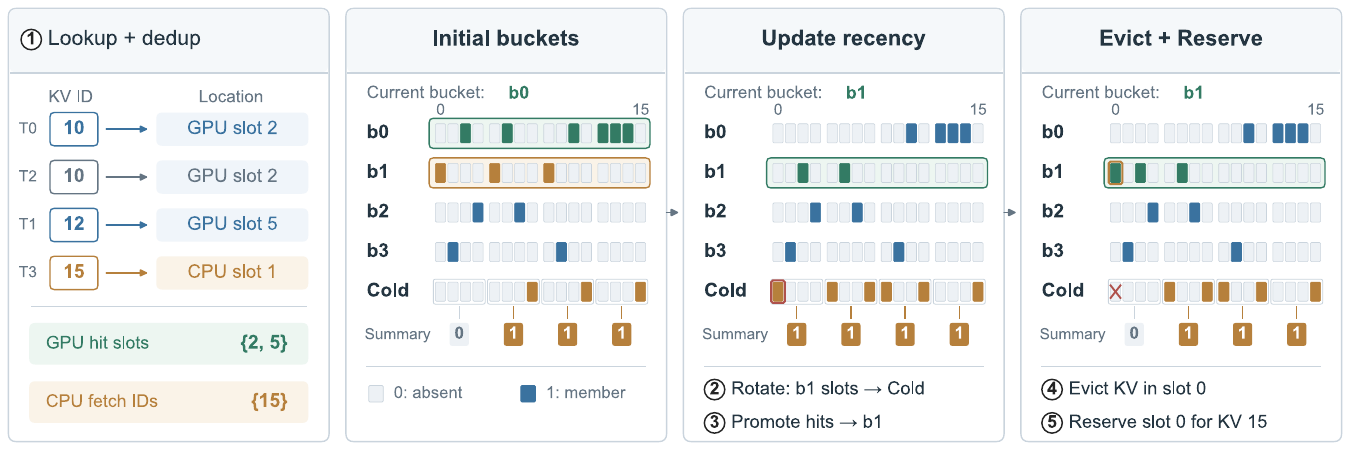}
\vspace{-10pt}
\caption{PB-LRU scheduling: (1) lookup and deduplication, (2) bucket rotation, (3) hit promotion, (4) GPU KV eviction, and (5) slot reservation.}
\vspace{-10pt}
\label{fig:pb-lru}
\end{figure*}

\subsection{KV Allocation and Access}
\label{sec:allocation-access}

Before prefill, the inference engine allocates logical IDs and records them in its request-to-token mapping. \sys\ assigns physical backing locally in CP and DPA or across storage ranks in TP. Prefill writes each layer's KV through these IDs; decoding retains them and allocates IDs for new tokens. Section~\ref{sec:api} describes the interface and write ordering.

The inference engine issues sparse access using logical IDs and a layer identity. In CP and DPA, \sys\ resolves these IDs to physical slots in the same rank's GPU or CPU pools. It promotes selected CPU data into that rank's GPU pool and supplies attention directly from the pool. In TP, \sys\ first reuses local buffer hits without consulting the storage address mapping, then resolves only misses to physical slots on their storage ranks. It promotes selected CPU data to the corresponding storage GPU before fetching missing KV into the computing rank's buffer. Attention reads after these transfers complete. Section~\ref{sec:residency} details promotion and mapping updates.

\subsection{Local GPU Buffer}
\label{sec:local-buffer}

Each TP computing rank caches selected KV in a local GPU buffer sized according to the configured request limit, independently of retained context length. Replacing a copy preserves the KV cache in the storage pools. CP and DPA read their local GPU pools directly without this buffer.

\textbf{Lookup and transfer.} Buffer tags identify cached logical addresses. Hits are protected for the current attention; misses are deduplicated across queries and assigned buffer slots. GPU kernels fetch from local pools or from peers through CUDA IPC into the attention backend's layout. In Figure~\ref{fig:gpu-pooling}, ID~2 is reused directly from the local buffer, bypassing address mapping. The misses for IDs $[4,5]$ resolve to GPU~0 slot~0 and GPU~1 slot~1; their KV is then read into GPU~0's local buffer.

\textbf{Plan reuse and prefetching.} Layers sharing sparse selections reuse address lookup, miss deduplication, and slot assignments, while fetching their own KV data. A separate CUDA stream overlaps prefetching with computation. Stream dependencies and events order source writes, transfers, and attention reads, preventing premature buffer reuse.

\subsection{Distributed Indexer}
\label{sec:index}

In TP, Index keys follow the KV cache's logical allocation and GPU partition. Each rank scores local keys, and an all-gather combines scores for the model's TopK selection. This reduces key replication and redundant scoring; selected positions enter the logical KV access path. The score exchange introduces minor communication and synchronization overhead. DPA scores local requests' keys without this cross-rank exchange.

\section{CPU Memory Expansion and PB-LRU}
\label{sec:residency}

PB-LRU schedules selected CPU KV into a GPU pool shared across requests, retaining recently accessed data and assigning destinations for misses. \sys\ executes the transfers and commits the resulting mappings.

\subsection{Parallel Bucketed LRU}
\label{sec:pb-lru}

Sparse selections exhibit temporal locality~\cite{hisparse}. PB-LRU approximates LRU by grouping GPU slots accessed in the same scheduling round, avoiding an ordering of individual accesses. Figure~\ref{fig:pb-lru} shows the five scheduling steps.

\noindent\textbf{Parallel access processing (Step~1).} GPU threads resolve logical addresses concurrently and deduplicate repeated references, identifying GPU hits and collecting CPU addresses into a fetch list. Each distinct GPU hit requires one recency update; each distinct CPU record requires one transfer.

\noindent\textbf{Recency buckets (Steps~2--3).} Layers sharing sparse selections share a ring of $B$ recent buckets (default: $255$) and one cold bucket; Figure~\ref{fig:pb-lru} illustrates four recent buckets. Step~2 advances the current bucket; before reuse, its remaining members move to the cold bucket. Step~3 moves GPU hits into the current bucket. These operations change metadata only, leaving KV on the GPU.

Each bucket is a bitmap with one bit per GPU physical slot. A set bit records membership; a per-slot bucket identifier lets an access clear its previous membership and set the current bit. Access updates touch only referenced slots. Rotation processes the expiring bucket's remaining members, potentially the entire pool in the worst case. CPU slots have no bucket bits.

\noindent\textbf{KV eviction and slot reservation (Steps~4--5).} Step~4 selects free GPU slots first, then evicts slots from the cold bucket and recent buckets in oldest-first order. Excluding the current bucket protects data needed by the current access. Each hierarchical summary bit indicates whether a region below contains candidates, allowing selection to skip empty regions. Step~5 reserves selected slots in the current bucket for incoming KV. Figure~\ref{fig:pb-lru} illustrates evicting KV from GPU slot~0 and reserving that slot for CPU-resident KV~15.

\subsection{KV Exchange and Mapping Updates}
\label{sec:kv-exchange}

The transfer plan from Steps~1--5 records each incoming logical address, its CPU source, GPU destination, and any replaced logical address. Planning updates bucket membership without changing KV data or address mappings.

After prior readers complete, GPU kernels exchange incoming data with the GPU evictee, reusing the incoming record's CPU slot for the outgoing record. An empty GPU destination needs only a CPU-to-GPU copy; its source slot is released after the mapping update. Once transfers complete, \sys\ commits the new locations and signals dependent attention reads or TP buffer fetches. Logical IDs remain unchanged throughout.

\section{Programming Interface}
\label{sec:api}

\begin{figure}[t]
\centering
\includegraphics[width=\columnwidth,trim=0 0.6854bp 0 0bp,clip]{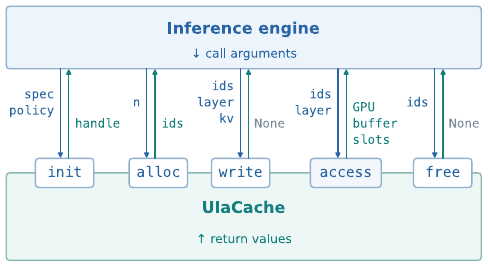}
\vspace{-10pt}
\caption{UlaCache API ports: each left arrow passes arguments downward; each right arrow returns results upward.}
\vspace{-10pt}
\label{fig:kv-api}
\end{figure}

Figure~\ref{fig:kv-api} shows the logical KV APIs exposed by \sys\ to the inference engine. The inference engine schedules requests and controls logical allocation, reuse, and release; \sys\ provides aggregate storage, physical allocation, address translation, residency, and KV transfers. This separation lets the engine combine \sys\ with features such as prefix caching and speculative decoding.

% Keep the accepted single-column API figure with Section 6; defer it to the following column.

\noindent\textbf{Initialization.} \texttt{init(spec, policy)} returns a cache handle. The specification defines per-layer KV record size and layout, scaling metadata, alignment, pool capacities, and Index-sharing groups. The policy binds execution resources and configures prefetching. Initialization creates pools, mappings, and workspaces.

\noindent\textbf{Allocation and reuse.} \texttt{alloc(n)} reserves backing for $n$ records and returns IDs from the bound allocator, including internal mapping registration. The inference engine stores these IDs in its request-to-token mapping and reuses them for shared prefixes.

\noindent\textbf{KV writes.} \texttt{write(ids, layer, kv)} writes native record segments or packed pages through allocated IDs. The prefill engine (P engine) populates prompt KV; the decoding engine uses the same operation for new tokens. \sys\ resolves physical destinations and preserves them until asynchronous writes complete.

\noindent\textbf{Sparse access.} \texttt{access(ids, layer)} prepares or reuses a residency plan, transfers required KV, and updates LRU state. It returns a borrowed \texttt{ResidentView}: GPU storage plus physical slot indices in selection order. Storage is the local rank's GPU pool in CP/DPA or the local buffer in TP. Logical IDs $[4,6,9]$, for example, may resolve to slots $[1,4,2]$. Physical indices can change across accesses while logical IDs remain stable. Layers sharing sparse selections reuse scheduling plans and enable later-layer prefetching.

\noindent\textbf{Release and capacity control.} The inference engine chooses slots for \texttt{free(ids)}. After their readers complete, \sys\ removes mappings, releases backing, and recycles IDs. Insufficient capacity makes \texttt{alloc} report failure; the inference engine decides what to release or whether to defer the request before retrying.

\section{Evaluation}
\label{sec:evaluation}

We compare HiCache~\cite{hicache}, HiSparse~\cite{hisparse}, and \sys\ in SGLang v0.5.22~\cite{sglang}, using GLM-5.3 on $8\times$H200 and HY4 Preview on $8\times$B300. All systems use FP8 KV cache under the memory budgets in Table~\ref{tab:platforms}. We evaluate DPA and TP. The speculative-decoding algorithm is multi-token prediction (MTP), with matching settings across systems: disabled (Spec. off) or two draft steps (Spec. on).

We use multi-turn agent traces from Finding the Right Fit~\cite{findingrightfit}, selecting sequence lengths of 256K--512K tokens for online serving tests. For throughput scaling, we use Multi-LexSum~\cite{multilexsum} inputs of 64K--128K tokens. We use long-context workloads to demonstrate the performance gains from KV capacity expansion at high concurrency.

\subsection{KV Allocation and Capacity}
\label{sec:evaluation-capacity}

\begin{figure}[!htb]
\centering
\includegraphics[width=\columnwidth,trim=0 -0.1648bp 0 1bp,clip]{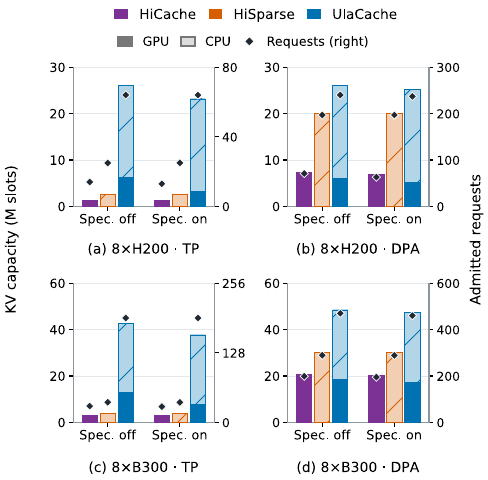}
\vspace{-10pt}
\caption{Effective KV capacity (bars, left axis) and admitted requests (diamonds, right axis).}
\vspace{-10pt}
\label{fig:ablation-capacity}
\end{figure}

Figure~\ref{fig:ablation-capacity} compares effective KV capacity and batch limits under matched memory budgets. The stacked capacity bars separate GPU and CPU storage, excluding replicas and buffers. \sys\ provides $9.3$--$11.4\times$ HiSparse's KV capacity in TP and $26$--$62\%$ more in DPA.

In DPA, \sys\ stores request KV in GPU and CPU memory. HiSparse's GPU buffers hold copies of its CPU-resident KV caches, adding no KV capacity. In TP mode, \sys\ also stores different portions of each KV cache on different ranks. For these capacity and throughput experiments, \sys's TP request limit is 64 on H200 and 192 on B300, bounding local GPU buffer memory rather than reflecting exhaustion of the KV pools. Enlarging the buffers admits more requests but leaves less GPU memory for KV storage.

\subsection{Throughput Scaling}
\label{sec:evaluation-throughput}

\begin{figure}[!htb]
\centering
\includegraphics[width=\columnwidth,trim=0 0bp 0 0bp,clip]{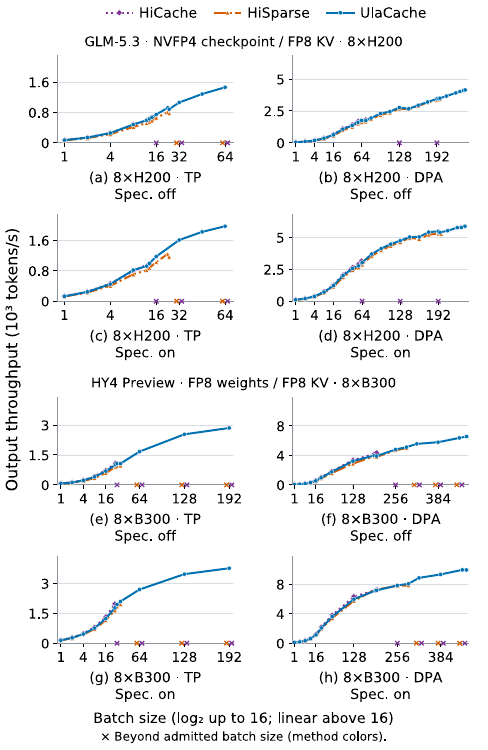}
\vspace{-10pt}
\caption{Decoding throughput across batch sizes.}
\vspace{-10pt}
\label{fig:evaluation_high_bs}
\end{figure}

We measure decoding throughput after prefill and compare each system's highest measured throughput. Figure~\ref{fig:evaluation_high_bs} shows output throughput across batch sizes; crosses mark batch sizes that cannot be admitted within the KV capacity and configured request limit. \sys's larger KV capacity keeps more requests active, extending throughput scaling beyond the baselines' admission limits. On $8\times$B300, its peak throughput in TP mode without speculative decoding is $2.62\times$ HiCache's peak and $2.95\times$ HiSparse's. The larger admissible batch lets more outputs share each batched model execution. The gain is larger in TP because KV replication consumes memory on multiple ranks. Partitioning this storage allows more of the aggregate GPU capacity to hold distinct KV.

With speculative decoding enabled in both systems, \sys\ achieves peak throughput speedups over HiSparse of $1.10\times$ (DPA) and $1.59\times$ (TP) on $8\times$H200, and $1.27\times$ (DPA) and $1.93\times$ (TP) on $8\times$B300. \sys's measured TP peaks occur at these limits. Speculative decoding further raises throughput at shared batch sizes by accepting multiple tokens per verification.

% Queue the online figure beside its discussion.
\begin{figure*}[!t]
\centering
\includegraphics[width=\textwidth,trim=0 0bp 0 0bp,clip]{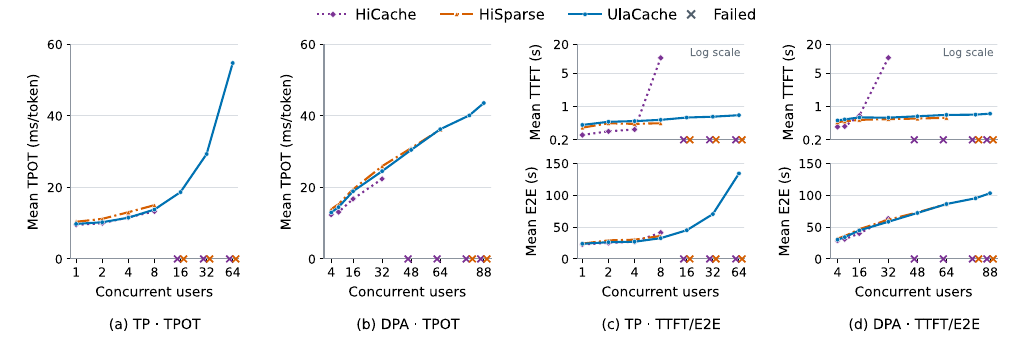}
\vspace{-10pt}
\caption{Agent-trace online serving on $8\times$H200 with speculative decoding. Crosses mark TTFT-limit failures.}
\vspace{-10pt}
\label{fig:evaluation-online}
\end{figure*}

\subsection{Online Serving for Agent Workloads}
\label{sec:evaluation-online}

We replay each conversation sequentially, reusing its KV cache across turns. An online run fails when the mean TTFT of its most recent 32 completed requests exceeds 120 seconds. Figure~\ref{fig:evaluation-online} compares TPOT, TTFT, and end-to-end latency as online concurrency increases.

\sys\ sustains $2.75\times$ HiCache's concurrency and $1.22\times$ HiSparse's (88 vs.\ 72 concurrent users) in DPA. In TP, it serves all tested concurrencies through 64 users, versus 8 for both baselines ($8\times$). HiCache limits active KV to GPU memory, while HiSparse stores complete KV in CPU memory and uses GPU space for cached copies. \sys\ stores distinct KV in both pools and, in TP, partitions KV across ranks, increasing the capacity available for concurrent conversations. Beyond the baselines' concurrency limits, \sys\ continues serving requests with increasing TPOT and end-to-end latency, while keeping mean TTFT below one second throughout the plotted range. At low concurrency, \sys\ achieves lower TPOT than HiSparse in both DPA and TP. As concurrency grows, HiCache's TTFT rises sharply, while \sys's TTFT remains stable. This difference can give \sys\ lower end-to-end latency than HiCache even when their TPOTs are comparable.

\subsection{Ablation and Overhead Analysis}
\label{sec:evaluation-ablation}

\begin{figure}[!htb]
\centering
\includegraphics[width=\columnwidth,trim=0 0.6103bp 0 0.5bp,clip]{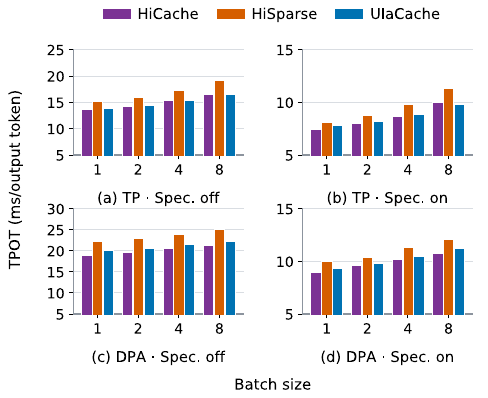}
\vspace{-10pt}
\caption{TPOT under low concurrency on $8\times$H200 for GLM-5.3.}
\vspace{-10pt}
\label{fig:evaluation_low_bs}
\end{figure}

\noindent\textbf{Latency under low concurrency.}\label{sec:evaluation-latency} Figure~\ref{fig:evaluation_low_bs} compares TPOT at low concurrency in DPA and TP. Across these configurations, \sys\ remains close to HiCache and consistently outperforms HiSparse. With speculative decoding, its TPOT overhead relative to HiCache ranges from $-2.4\%$ to $4.6\%$ in TP and from $2.6\%$ to $4.5\%$ in DPA. HiSparse incurs larger overheads of $9.3\%$--$14.0\%$ in TP and $8.6\%$--$11.9\%$ in DPA.

These results highlight the benefit of sharing GPU cache capacity when few requests are active. HiSparse limits each request to a fixed GPU buffer, leaving selected KV outside that buffer to be fetched from CPU memory even when other GPU cache space is idle. \sys\ lets active requests use this shared space to retain more KV, reducing CPU-to-GPU transfers. It therefore combines the additional capacity of CPU storage with TPOT close to the GPU-resident HiCache reference at low concurrency.

\noindent\textbf{PB-LRU ablation.}\label{sec:evaluation-scheduler} We compare HiSparse, a shared-pool HiSparse variant, a serial linked-list LRU, and PB-LRU by replaying the same GLM-5.3 DPA traces on H200 with speculative decoding enabled. This variant adapts HiSparse's original LRU kernel to a common pool. Shared policies use identical capacities and initial residency, while HiSparse retains fixed request buffers under the same total GPU budget. Timing covers GPU metadata only, excluding KV transfers.

\begin{figure}[!htb]
\centering
\includegraphics[width=\columnwidth,trim=0 1.875bp 0 0bp,clip]{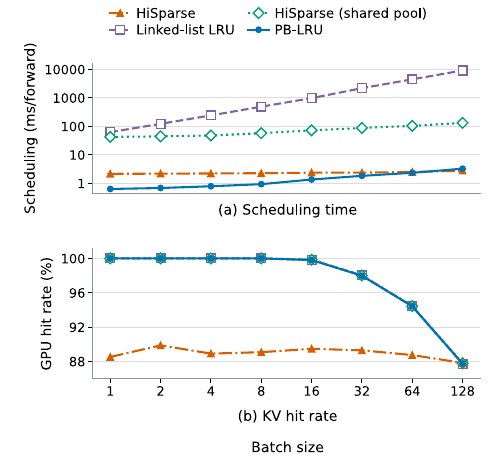}
\vspace{-10pt}
\caption{PB-LRU ablation on H200 (Spec. on): (a) GPU scheduling time; (b) KV hit rate.}
\vspace{-10pt}
\label{fig:ablation-scheduler}
\label{fig:ablation-schedule}
\label{fig:ablation-hit}
\end{figure}

Figure~\ref{fig:ablation-scheduler}(a) shows that PB-LRU schedules $40.6$--$67.5\times$ faster than the shared-pool HiSparse variant. Bucket bitmaps let PB-LRU update accessed slots and select evictees in parallel, whereas the adapted kernel scans and compacts the larger pool. HiSparse's smaller request buffers keep its scheduling cost low, and it becomes faster than PB-LRU at the largest batch size.

Figure~\ref{fig:ablation-scheduler}(b) shows that PB-LRU preserves hit rates close to the other shared-pool policies. At low concurrency, sharing lets active requests borrow idle capacity and retain their complete KV cache, improving hit rates over HiSparse. As concurrency grows, the pool fills and this advantage narrows.

The shared-pool comparison shows why allocation and replacement must be considered together. Pooling capacity improves reuse, but adapting a kernel designed for small request buffers can make metadata processing expensive. PB-LRU retains the hit-rate benefit of pooling while avoiding a serial ordering of slot accesses.

\begin{figure}[!htb]
\centering
\includegraphics[width=\columnwidth,trim=0 -1.8273bp 0 0bp,clip]{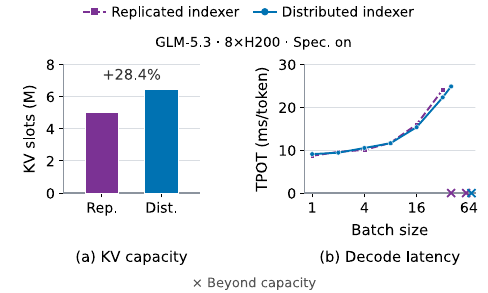}
\vspace{-10pt}
\caption{Indexer ablation on $8\times$H200 (Spec. on), with all KV and Index keys in GPU memory.}
\vspace{-10pt}
\label{fig:ablation-indexer}
\end{figure}

\noindent\textbf{Distributed indexer ablation.}\label{sec:evaluation-indexer} We compare replicated and distributed indexers for GLM-5.3 using TP on $8\times$H200 with speculative decoding. Both configurations partition the target KV cache across GPUs; all KV and Index keys remain in GPU memory. Replicated indexing stores a full copy of Index keys on every rank; distributed indexing partitions these keys as described in Section~\ref{sec:index}.

In Figure~\ref{fig:ablation-indexer}(a), the distributed configuration provides $28.4\%$ more usable KV capacity under the same GPU memory budget. Partitioning Index keys leaves more GPU memory for KV storage. Figure~\ref{fig:ablation-indexer}(b) shows similar TPOT at low concurrency, with lower TPOT for the distributed configuration at the larger common batch sizes. Thus, distributing Index keys frees additional capacity even after target KV has been partitioned.

\section{Related Work}
\label{sec:related}

\paragraph{KV cache allocation.}
PagedAttention uses block allocation to reduce GPU memory fragmentation~\cite{pagedattention}. vAttention and vTensor use GPU virtual memory management to separate address layout from physical allocation~\cite{vattention,vtensor}. \sys\ extends logical KV addressing across GPU and CPU pools, allowing both to contribute capacity while preserving references as KV moves between them.

\paragraph{Prefix caching.}
Prompt Cache reuses predefined prompt modules~\cite{promptcache}, while SGLang organizes shared prefixes in a radix tree~\cite{sglang}. CachedAttention, HiCache, and LMCache retain reusable KV beyond GPU memory~\cite{cachedattention,hicache,lmcache}; Mooncake extends KV reuse through distributed storage~\cite{mooncake}. These systems save prefill computation by restoring previously computed KV. \sys\ complements reuse by expanding the KV capacity available to concurrent requests: the engine decides which prefixes to retain, while \sys\ manages their storage and supplies KV for attention.

\paragraph{KV cache offloading.}
FlexGen offloads tensors to CPU memory and disk when GPU memory is limited~\cite{flexgen}. For sparse attention, HiSparse keeps complete KV in CPU memory with fixed per-request GPU buffers~\cite{hisparse}; SPIN supports elastic buffers and GPU bucketed LRU~\cite{spin}. Their GPU buffers cache CPU-backed copies. \sys\ lets GPU and CPU pools store different KV and uses PB-LRU to manage a GPU pool shared across requests. We compare directly with HiSparse in the same engine; no public SPIN implementation was available for our evaluation.

ECHO overlaps KV recall with indexer computation through lossless prefetching~\cite{echo}. vLLM's Hybrid HiSparse retains KV on GPUs until memory pressure requires offloading, sharing GPU pages between resident KV and hot buffers~\cite{vllmhybridhisparse}. \sys\ could adopt these prefetching and offloading techniques to reduce KV transfer overhead while using the combined GPU and CPU capacity to support more concurrent requests.

\paragraph{Sparse attention.}
Quest selects KV pages using query-dependent scores~\cite{quest}. MInference accelerates prefill with structured sparsity~\cite{minference}, and LServe combines static and dynamic sparsity in prefill and decoding~\cite{lserve}. InfiniGen, RetrievalAttention, and OasisKV reduce KV transfers through prediction, retrieval, and lookahead selection~\cite{infinigen,retrievalattention,oasiskv}. FlashInfer provides customizable attention kernels~\cite{flashinfer}, while SparseEngine integrates diverse sparse methods~\cite{sparseengine}. \sys\ accepts the model's existing selections and manages the selected KV through a common memory interface, enabling storage optimization without changing the attention algorithm.

\paragraph{Parallel inference.}
Infinite-LLM pools GPU memory and distributes attention across devices~\cite{infinitellm}. Helix combines KV parallelism for attention with TP for feed-forward computation~\cite{helix}. \sys\ provides KV access paths for existing CP, DPA, and TP layouts over GPU and CPU storage. In TP configurations with replicated KV, it stores different portions across ranks and gathers selected KV for local attention, reducing storage redundancy while preserving the engine's attention computation.

\section{Conclusion}
\label{sec:conclusion}

\sys\ combines unified GPU--CPU KV addressing, access paths adapted to attention parallelism, and PB-LRU cache management to increase usable capacity and reduce sparse-access overhead. Under matched memory budgets, \sys\ achieves $1.10\times$ (DPA) and $1.59\times$ (TP) the peak decoding throughput of HiSparse on $8\times$H200, and $1.27\times$ (DPA) and $1.93\times$ (TP) on $8\times$B300, with speculative decoding enabled in both systems.

\bibliographystyle{mlsys2025}
\bibliography{references}

\end{document}